\documentclass[bibyear]{aa} 

\usepackage{graphicx}
\usepackage{txfonts}
\begin{document}
   \title{Flare in the Galactic stellar outer disc detected in SDSS-SEGUE data}

   \subtitle{}

   \author{M. L\'opez-Corredoira\inst{1,2}, J. Molg\'o\inst{3}}
\institute{
$^1$ Instituto de Astrof\'\i sica de Canarias,
E-38205 La Laguna, Tenerife, Spain\\
$^2$ Departamento de Astrof\'\i sica, Universidad de La Laguna,
E-38206 La Laguna, Tenerife, Spain\\
$^3$ GRANTECAN S.A., E-38712, Breña Baja, La Palma, Spain
}

\offprints{martinlc@iac.es}
\titlerunning{Flare}
\authorrunning{L\'opez-Corredoira}

   \date{Received xxxx; accepted xxxx}

 
  \abstract
    {}  
  {We explore the outer Galactic disc up to a Galactocentric distance of $\approx $30 kpc to derive its parameters and measure the magnitude of its flare.}
  {We obtained the 3D density of stars of type F8V-G5V with a colour selection from
extinction-corrected photometric data of the Sloan Digital Sky Survey -- Sloan Extension for Galactic Understanding and Exploration (SDSS-SEGUE) over 1,400 deg$^2$ in off-plane low Galactic latitude regions and fitted it to a model of flared thin+thick disc.}
  {The best-fit parameters are a thin-disc scale length of 2.0 kpc, a thin-disc scale height at solar Galactocentric distance of 0.24 kpc, a thick-disc scale length of 2.5 kpc, and a thick-disc scale height at solar Galactocentric distance of 0.71 kpc. We derive a flaring in both discs that causes the scale height of the average disc to be multiplied with respect to the solar neighbourhood value by a factor of $3.3^{+2.2}_{-1.6}$ at $R=15$ kpc and by a factor of $12^{+20}_{-7}$ at $R=25$ kpc.}
  {The flare is quite prominent at large $R$ and its presence explains the apparent depletion of in-plane stars that are often confused with a cut-off at $R\gtrsim 15$ kpc. Indeed, our Galactic disc does not present a 
truncation or abrupt fall-off there, but the stars are spread in off-plane regions, even at $z$ of several kpc for $R\gtrsim 20$ kpc.
Moreover, the smoothness of the observed stellar distribution also suggests that there is a continuous structure and not a combination of a Galactic disc plus some other substructure or extragalactic component: the hypothesis to interpret the Monoceros ring in terms of a tidal stream of a putative accreted dwarf galaxy
is not only unnecessary because the observed flare explains the overdensity in the Monoceros ring observed in SDSS fields, but it appears to be inappropriate.}

   \keywords{Galaxy: structure --- Galaxy: disc --- Galaxy: stellar content}

   \maketitle
%

\section{Introduction}

We know many things about the structure of our Galaxy, and we have an approximate idea
of the functional shape of its components: thin and thick disc, bulge, long bar, halo
or spheroid, spiral arms, ring. In the past decades, different large-area surveys in visible
or near-infrared wavelengths have allowed us to know our Galaxy much better.
However there are some parts of the Galaxy that are not well known; one of them is the
outer part of the disc, with Galactocentric distances beyond $R=15$ kpc.
We know this component is warped and flared, but the
details of its shape are still a topic to develop further.

We are interested in analyse this outer part of stellar disc in our Galaxy in this paper,
using available visible data of the Sloan Digital Sky Survey (SDSS) in low Galactic latitude regions.
Several authors (Bilir et al. 2006, 2008; Juri\'c et al. 2008; Jia et al. 2014) 
have previously used the SDSS to measure the parameters of the disc, but they have explored the high Galactic latitudes, so they could not access the outer disc. Other authors have analyzed the disc 
using the Two Micron All Sky Survey (2MASS) either at low latitudes 
(L\'opez-Corredoira et al. 2002, 2004) or at high latitudes using red-clump giants (Cabrera-Lavers et al. 2005, 2007; Chang et al. 2011) or within the whole sky using a given luminosity function (Polido et al. 2013), but, even using near-plane regions, the depth of 2MASS is not enough to analyze the
outer disc; it is dominated by stars with smaller distances than the outer disc we
would like to analyze.

Our purpose here is to use a survey like the SDSS (\S \ref{.data}), deeper than 2MASS, 
in near-plane regions, although avoiding the plane itself because of the high extinction. From this we separate a type of dwarf stars as
standard candles (\S \ref{.method}).
We are particularly interested in modelling the flare of the outer disc. There
is some evidence of its existence in the different components of the Galaxy, but
the stellar flare at $R\gtrsim 15$ kpc is still poorly known.
The flare of HI was investigated for instance by Nakanishi \& Sofue (2003), Levine et al. (2006), or Kalberla et al. (2007). It was modelled by Kalberla et al. (2007) in terms of a dark matter ring, by Saha et al. (2009) in terms of a lopsided halo,
or by L\'opez-Corredoira \& Betancort-Rijo (2009) in terms of accretion of intergalactic matter onto the disc.
Another example of modelling is given by Narayan \& Jog (2002), who
used a three-component (HI, H$_2$ and stars) disc gravitationally coupled to calculate the scale height of each component and their flares, and they derived a good prediction for the gas flare;
the result of mild stellar flaring up to $R=12$ kpc was obtained in the model by Narayan \& Jog (2002), and it agreed well with the  observational data on flaring provided by Kent et al. (1991). 
The stellar flare was also observed by
Alard (2000), L\'opez-Corredoira et al. (2002), Yusifov (2004), Momany et al. (2006), or Reyl\'e et al. (2009) for the outer disc, but limited to $R\lesssim 20$ kpc and with large uncertainties over 15 kpc or for the differences between the flare of the thin and thick disc.

A flare in the thick disc is 
posited (Hammersley \& L\'opez-Corredoira 2011; Mateu et al. 2011; L\'opez-Corredoira et al. 2012), 
and this may constitute an explanation for the Monoceros ring instead of the tidal stream
hypothesis. This is an additional reason to pursue this research: to know whether we are able
to fit our star counts without needing some new extragalactic component. The fact that
a flared thin+thick disc alone fits our data (\S \ref{.fit}, \S \ref{.explor}) gives a positive answer,
which is discussed and compared with other works of the literature in \S \ref{.compar}, \S \ref{.discu}, together with the considerations derived from the morphological information obtained in this paper.

\section{Data} 
\label{.data}

The SDSS (Sloan Digital Sky Survey)-DR8 release (Aihara et al. 2011) contains imaging of 
14\,555 deg$^2$ of the sky in five filters (u,g,r,i,z), with a completeness higher than 95\% 
for $m_g\le 22.2$. Within this survey, the subsample SEGUE (Sloan Extension for Galactic Understanding 
and Exploration) includes many Galactic plane regions.
For this paper, in which we are interested to explore the outer disc, we used the regions with 
$|\ell |> 50^\circ $ (to avoid the regions of the inner Galaxy), 
$|b|\le 23^\circ $ (dominated by disc stars), taking all the point-like SDSS sources with
the flags and constraints for the $g$ and $r$ filters: ((flags\_r\,\& \,0x10000000)\, != 0),
((flags\_r\, \& \,0x8100000c00a4)\, = 0), (((flags\_r\, \& \,0x400000000000)\, = 0)\, or\, 
(psfmagerr\_r <= 0.2)),
(((flags\_r\, \& \,0x100000000000)\, = 0)\, or\, (flags\_r\, \& \,0x1000)\, = 0), plus the same thing for 
flags\_g;
this avoids the stars that are too close to the borders, are too small to determine
the radial profile, have saturated pixels or too many interpolated pixels
to derive a correct flux, stars for which the deblend algorithm finds two or more
candidates in cases with a magnitude error in $r$ larger than 0.2, stars with poor detection, or cases with cosmic rays.
All the magnitudes were corrected for extinction by the SDSS team using the 
Galactic extinction model of Schlegel et al. (1998). 
In total, we cover an area of 1\,745 deg$^2$.
As we explain in the next section, we reduced this area by avoiding the regions very close to the plane with high extinction to reduce the errors due to the correction of extinction.

\section{Method}
\label{.method}

The simplest method of determining the stellar density along a 
line of sight in the disc is by isolating a group of stars with the same colour and absolute magnitude 
$M$ within a colour magnitude diagram. This allows the luminosity function to
be replaced by a constant in the stellar statistics equation and the differential star counts for each line of sight, $A(m)$, can be immediately converted into density $\rho (r)$:

\begin{equation}
\label{diffsc}
A(m)\equiv \frac{dN(m)}{dm}=\frac{ln\ 10}{5}\omega \rho
[r(m)]r(m)^3  
,\end{equation}\[
r(m)=10^{[m-M+5]/5}
,\]
where $\omega $ is the area of the solid angle in radians and $r$ is the
distance in parsecs.

In the near-infrared, red-clump giants have been successfully used as  standard candles, particularly for the innermost 15 kpc from the Galactic centre
(L\'opez-Corredoira et al. 2002). However, red-clump stars in the outer disc at the distance of interest would appear at $m_k\ge 14$, where the local dwarfs with the same colour would completely dominate the counts 
(L\'opez-Corredoira et al. 2002). Therefore, we have to use something different here.

For our extinction-corrected areas, it is possible to use star counts in visible. An examination of the HR diagram shows that when the extinction is low, the late F and early G dwarfs can be isolated using colour with only minimal contamination from other sources with the same colour, but different
absolute magnitudes. For this work we selected the 
sources between F8V and G5V. 
There will be sufficient stars detected in the outer Galaxy to give meaningful statistics, 
which would not be the case if a smaller range in absolute magnitudes were used. Earlier sources were
not included as these sources would belong to a younger population with a 
far lower scale height; the absolute magnitude also changes
far more rapidly with colour. Later sources would have significant giant contamination, 
and again the absolute magnitude changes more rapidly with colour. 

   \begin{figure}
\vspace{1cm}
   \centering
   \includegraphics[width=8cm]{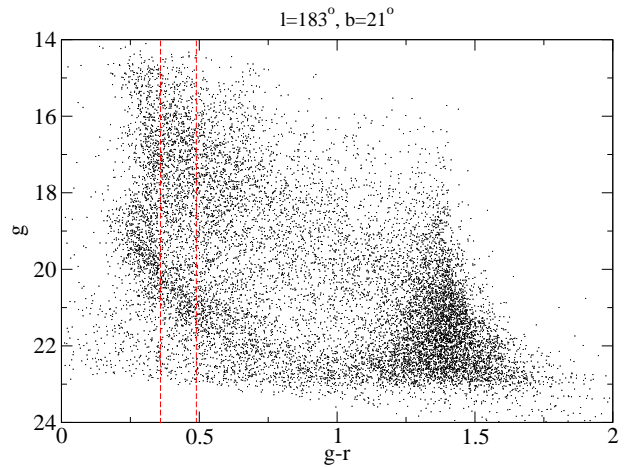}
   \caption{Extinction-corrected colour-magnitude diagram for the region $\ell=183^\circ $, $b=21^\circ$. The region between the dashed lines contains F8V-G5V dwarfs.}
   \label{Fig:CMdiagram}
   \end{figure}

The F8V-G5V dwarfs have a range of $g-r$ of 0.36 to 0.49 
and a range of absolute magnitudes $M_g$=4.2 to 5.3 (Bilir et al. 2009),
which makes the sources approximately $m_g\lesssim 21.5$ 
at the distance of interest up to 22 kpc. See the selection example in Fig. \ref{Fig:CMdiagram}. We adopted a constant absolute magnitude for all of them
$M_g$=4.8. For a range of absolute magnitudes, when the counts are converted into 
density vs. distance, there is some smoothing which is not included in a model that assumes a single absolute magnitude. This smoothing has little effect,  
and the above approximation remains valid: see L\'opez-Corredoira et al. 2002, \S 3.3.1
for a calculation of the difference between a narrow Gaussian distribution and a Dirac delta: it leads to an error in the scale length of the order of 2\% assuming an r.m.s. in the Gaussian distribution of 0.3 mag.; although the application of L\'opez-Corredoira et al. is for red-clump giants, 
it is valid for any kind of population.
We did not take into account the possibility that some of these stars might indeed be binary systems.
See Siegel et al. (2002) or Bilir et al. (2009, \S 3.4) for a calculation of 
the effects of a high ratio of
binary stars to derive the parameters of the disc: they may produce 
$\Delta M_g\sim 0.2$ mag.
There are some radial and vertical gradients of metallicity for
thin and thick disc (Rong et al. 2001; Ak et al. 2007; Andreuzzi et al. 2011); 
here, like in Juri\'c et al. (2008),
we did not take into account the variation of the absolute magnitude due to the variation of metallicity: 
see Siegel et al. (2002) or Juri\'c et al. (2008, \S 2.2.1) for a discussion on it. 
A rough estimate of the maximum difference for the most extreme cases of 
lower metallicity for the highest values of $R$ and $z$ ($[Fe/H]\sim 1$ dex
lower than in the solar neighbourhood) is $\Delta M_R\approx 0.4$ [Siegel et al. 2002, using $R-I=0.38$, which is the corresponding transformation from SDSS to Johnson filters (Jordi et al. 2006) of the
color of our population with average $(r-i)=0.13$ (Bilir et al. 2009)]. The variation of $M_g$ is
similar (Juri\'c et al. 2008, Fig. 3). Therefore, 
the fact that we did not take this metallicity gradient into account 
may produce an overestimate of
the distance of the stars at the highest $R$ or $z$ of $\sim 20$\% . Its effects are explored in \S 
\ref{.explor}.
 
\begin{figure}
\vspace{1cm}
\centering
\includegraphics[width=8.5cm]{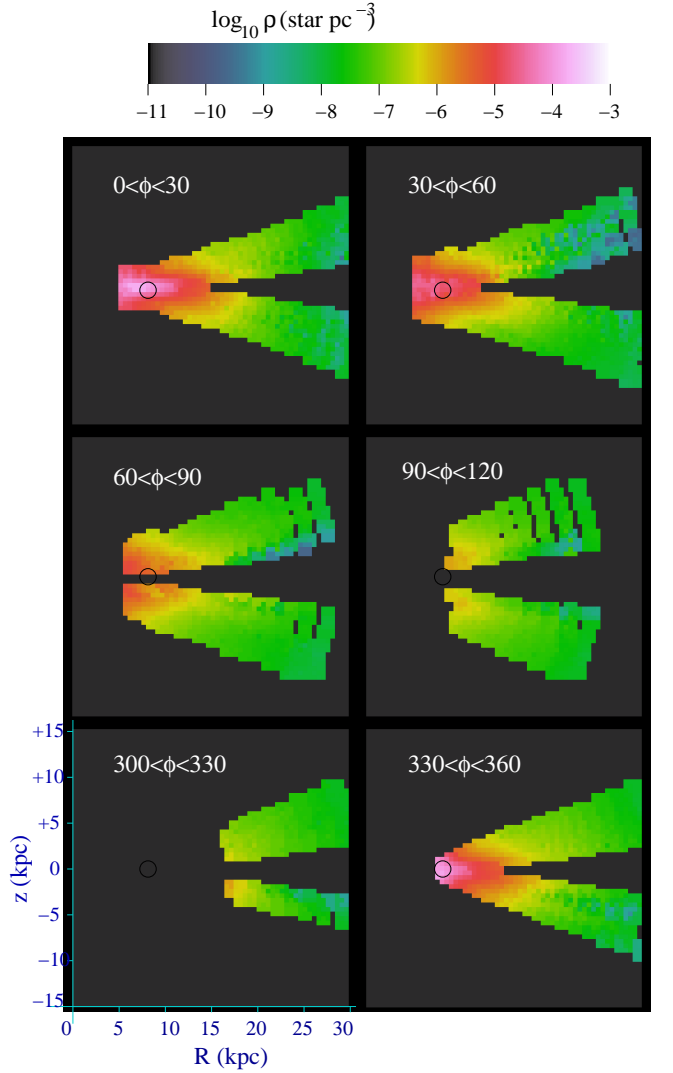}
\caption{Average stellar density of F8V-G5V stars from available SDSS data as a function of the 
cylindrical coordinates $R$ (kpc), $\phi $ (deg.), $z$ (kpc)
for the range $0<R<30$ kpc, $|z|<15$ kpc, constrained within Galactic latitudes
$8^\circ \le |b|\le 22^\circ $. The circle stands for $R=8$ kpc and $z=0$ (like the
Sun position) for any value of $\phi $ (for the Sun, it would be $\phi =0 $).
The axes in the bottom left panel apply to all panels in the figure.}
\label{Fig:dens}
\end{figure}

In the 1\,745 deg$^2$ of our SDSS data are 6\,506\,360 stars with $g-r$ (corrected for extinction) between 0.36 and 0.49. The source densities in these regions are expected to be very low, therefore we require regions of more than 0.5 square degrees of sky to be covered to provide sufficient counts to give reasonable statistics. 
We divided them into multiple space bins with $\Delta \ell \times \Delta b=2^\circ \times 2^\circ $, $\Delta m_g=0.2$ for $15.25<m_g<21.85$, and,
since we know their distances and coordinates (hence, we know their position in 3D space), given the differential star counts $A(m)$, we can derive through Eq. (\ref{diffsc}) the density $\rho (R,\phi ,z)$ 
(in cylindrical Galactocentric coordinates). This average stellar density is plotted in Fig.
\ref{Fig:dens}, excluding the bins with $|b|<8^\circ $.

We excluded these in-plane regions (bins with $|b|<8^\circ $), which reduced 
our area to 1\,396 deg$^2$, to reduce the errors
in the extinction correction: with extinctions of $\langle A(g)\rangle \lesssim 1.5$ mag 
in filter g at most for $|b|\ge 8^\circ $ (Schlegel et al. 1998), $E(g-r)=0.27A(g)$ 
and a relative uncertainty of the reddening of 
$\sim 10$\% (Schlegel et al. 1998; M\"ortsell 2013), we derive relative
errors of $\Delta (g-r) \lesssim 0.04$ for each region. With a $\frac{dM_g}{d(g-r)}\approx
\frac{5.4-4.2}{0.49-0.36}=9.2$, we compute uncertainties in the distance determination
of $\lesssim 18$\%. Although this error is still large, when combining all the bins,
the errors compensate for each other and are mostly cancelled on except for some systematic
ones that may lead to some variation of the scales. We discuss this point in \S \ref{.discu}. At present, we do not carry out a deconvolve the line of sight distribution with the spread of distances because of the r.m.s. of the reddening, which would provide a better determination of the morphology, because one would need a very accurate expression of the r.m.s. of the extinction, which is not available (see also discussion in \S \ref{.discu}). Because the regions are all well off the plane we assumed that 
this extinction is local.

\section{Fitting free parameters of a disc model}
\label{.fit}

\subsection{Disc model}

After deriving the stellar density $\rho (R,\phi ,z)$, we can
fit a disc model to obtain its best free parameters. We
used the following model, which contains flared axysymmetric thin+thick discs:
             
\begin{equation}
\label{rho}
\rho _{\rm disc}(R,z)=\rho _{\rm thin}(R,z)+\rho _{\rm thick}(R,z)
,\end{equation}
\[
\rho_{\rm thin} (R,z)= \rho_\odot \frac{h_{\rm z, thin}(R_\odot)}{h_{\rm z, thin}(R)}\exp\left( \frac{R_\odot}{h_{\rm r, thin}} + \frac{h_{\rm r, hole}}{R_\odot} \right)
\]\[\ \ \ \ \times 
\exp\left( -\frac{R}{h_{\rm r, thin}}-\frac{h_{\rm r, hole}}{R} \right) \exp\left( -\frac{|z|}{h_{\rm z, thin}(R)} \right)
,\]\[
\rho_{\rm thick} (R,z)= f_{\rm thick}\,\rho_\odot \frac{h_{\rm z, thick}}{h_{\rm z, thick}(R)}\exp \left( \frac{R_\odot}{h_{\rm r, thick}} + \frac{h_{\rm r, hole}}{R_\odot} \right)
\]\[\ \ \ \ \times \exp \left( -\frac{R}{h_{\rm r, thick}}-\frac{h_{\rm r, hole}}{R} \right) \exp
\left( -\frac{|z|}{h_{\rm z, thick}(R)} \right)
,\]
where the respective scale heights are
\begin{equation}
h_{\rm z, thin}(R) = h_{\rm z, thin}(R_\odot)\left( 1 + \sum_{i=1}^2 k_{\rm i, thin}(R-R_\odot )^i \right)
,\end{equation}\[
h_{\rm z, thick}(R) = \left\{ \begin{array}{lcl}
      h_{\rm z, thick}(R_{\rm ft})&  R<R_{\rm ft}\\
      h_{\rm z, thick}(R_{\rm ft})\left( 1 + \sum_{i=1}^2 k_{\rm i, thick}(R-R_{\rm ft})^i\right) & R \geq R_{\rm ft}\\
   \end{array}
   \right.
.\]
This expresses an exponentially flared disc (L\'opez-Corredoira et al. 2002) with a hole or deficit of
stars in the inner in-plane region (L\'opez-Corredoira et al. 2004). Since we did not examine the inner
disc, we kept $h_{\rm r, hole}$ constant: $h_{\rm r, hole}=3.74$ kpc (L\'opez-Corredoira et al. 2004).
We also kept constant the ratio of thick to thin disc stars in the solar neighbourhood,
$f_{\rm thick}=0.09$; and the rest of the parameters were left free in our fit: 
the scale length of the thin disc $h_{\rm r, thin}$, the scale height of the thin disc at solar Galactocentric
distance $h_{\rm z, thin}(R_\odot)$, the scale length of the thick disc $h_{\rm r, thick}$,
the scale height of the thin disc at solar Galactocentric
distance $h_{\rm z, thick}(R_\odot)$, the two parameters defining the flare of the thin disc
$k_{\rm 1, thin}$ and $k_{\rm 2, thin}$, the two parameters defining the flare of the thick disc $k_{\rm 1, thick}$, $k_{\rm 2, thick}$, and the scale at which the thick disc flare starts, $R_{\rm ft}$.

Bovy et al. (2012) suggested that there is no thick disc that sensibly can be 
characterized as a distinct component
because mono-abundance sub-populations, defined in the [alpha/Fe]-[Fe/H] space, are well described by single-exponential spatial-density profiles in both the radial and the vertical direction; therefore, any star of a given abundance would be associated with a sub-population of a given scale height and there would be a continuous and monotonic distribution of disc thicknesses. This result is contradicted by other authors however (e.g., Haywood et al. 2013), who showed a bimodal distribution in the [alpha/Fe]-[Fe/H] space, or that thin and thick disc can be distinguished because of the rotation and vertical dispersion, age, or other indicators of metallicity. Here we assumed that there are morphologically two discs and do not engage in the discussion about their populations.

We also included a halo, but did not try to fit any of its parameters; we just took
the fixed density distribution provided by Bilir et al. (2008) using SDSS data. 
Its contribution to our counts is lower than the disc contribution (20\% for the highest $R$ and $z$ of
our range).
\begin{equation}
\rho_{\rm halo} (R,z)= 1.4\times 10^{-3}\,\rho_\odot \frac{\exp\left[10.093\left( 1 - \left(\frac{R_{\rm sp}}{R_\odot} \right)^{1/4} \right) \right]}{(R_{\rm sp}/R_\odot )^{7/8}} 
,\end{equation}\[
R_{\rm sp} = \sqrt{R^2+2.52z^2}
.\]

Our model does not take the stellar warp into account.
The warp moves the position of the Galactic plane: the plane can be 
shifted away from the expected position of $b=0$. It can be clearly
seen in the counts within a few kpc of the Sun (L\'opez-Corredoira et al. 2002, 
Momany et al. 2006, Reyl\'e et al. 2009) and its effect is stronger towards 
the outer edge of the disc.
The models of the warp are normally simple, based on tilted rings, and in general 
represent the star count well. Nonetheless, its average effect is not important for
$|b|\ge 8^\circ $ (see Fig. 15/top of L\'opez-Corredoira et al. 2002 with ratio of positive/negative counts
close to unity): some regions are overdense in the positive latitudes with respect to the negative latitude for the northern warp region (the southern warp is not visible with SDSS), but on average for our fit both positive and negative latitude data cancel each other out and only a 
minor second-order effect may remain in the fit, which we neglected. 
In Fig. \ref{Fig:dens}, no significant signs of the northern warp are detected
(the southern warp is not visible in our SDSS data). A more detailed 
discussion on the influence of the warp is given in the
last paragraph of the next subsection and in \S \ref{.explor}.

\subsection{Best fit}

\begin{figure*}
\vspace{1cm}
\centering
\includegraphics[width=18cm]{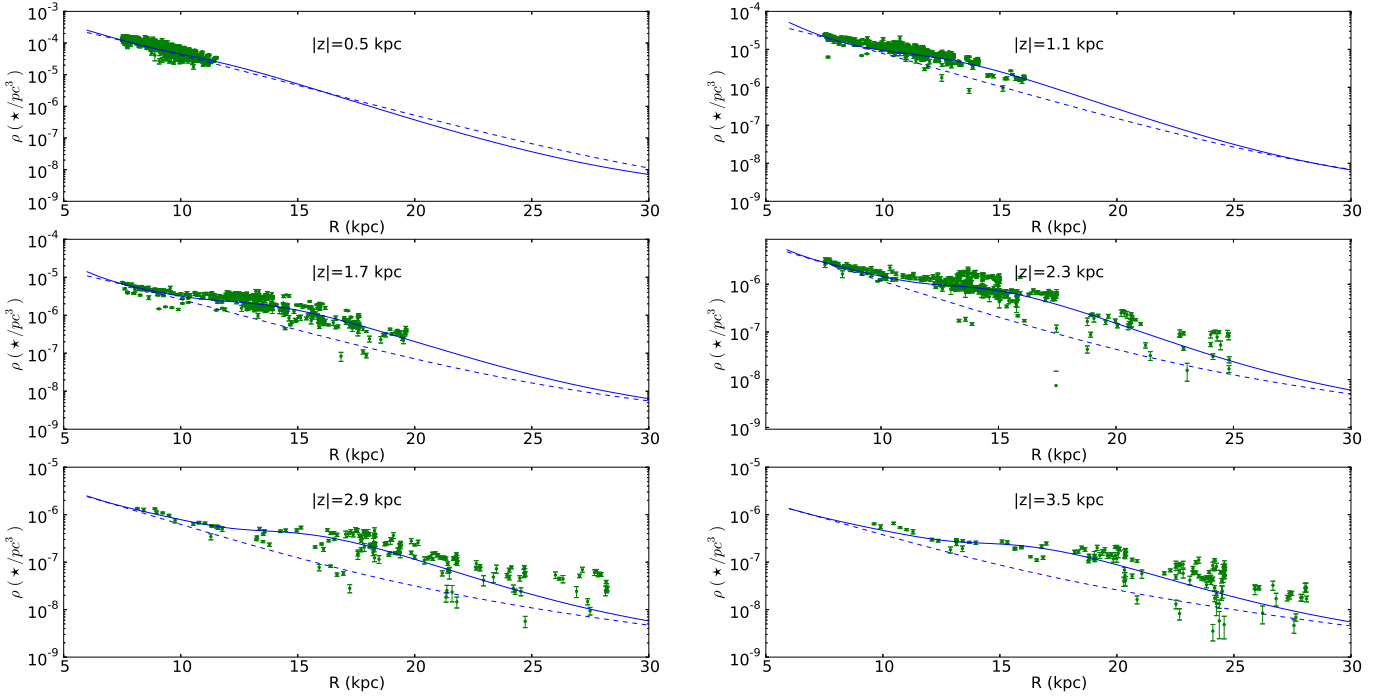}
\caption{Best fit (solid line) for the data of $\rho (R,z)$ extracted from the star counts of
F8-G5V stars in the SDSS. The dashed line stands for the best fit of the disc within $R<15$ kpc
without any flare, and extrapolated for $R\ge 15$ kpc.
Error bars stand only for Poissonian errors in the star counts and
do not include other possible factors such as errors in the extinction.
Note that in the lowest $|z|$ bins there are no 
points at high $R$ because of our constraint of 
$|b|>8^\circ $ (see Fig. \ref{Fig:dens}).}
\label{Fig:fitres}
\end{figure*}

\begin{figure}
\vspace{1cm}
\centering
\includegraphics[width=9cm]{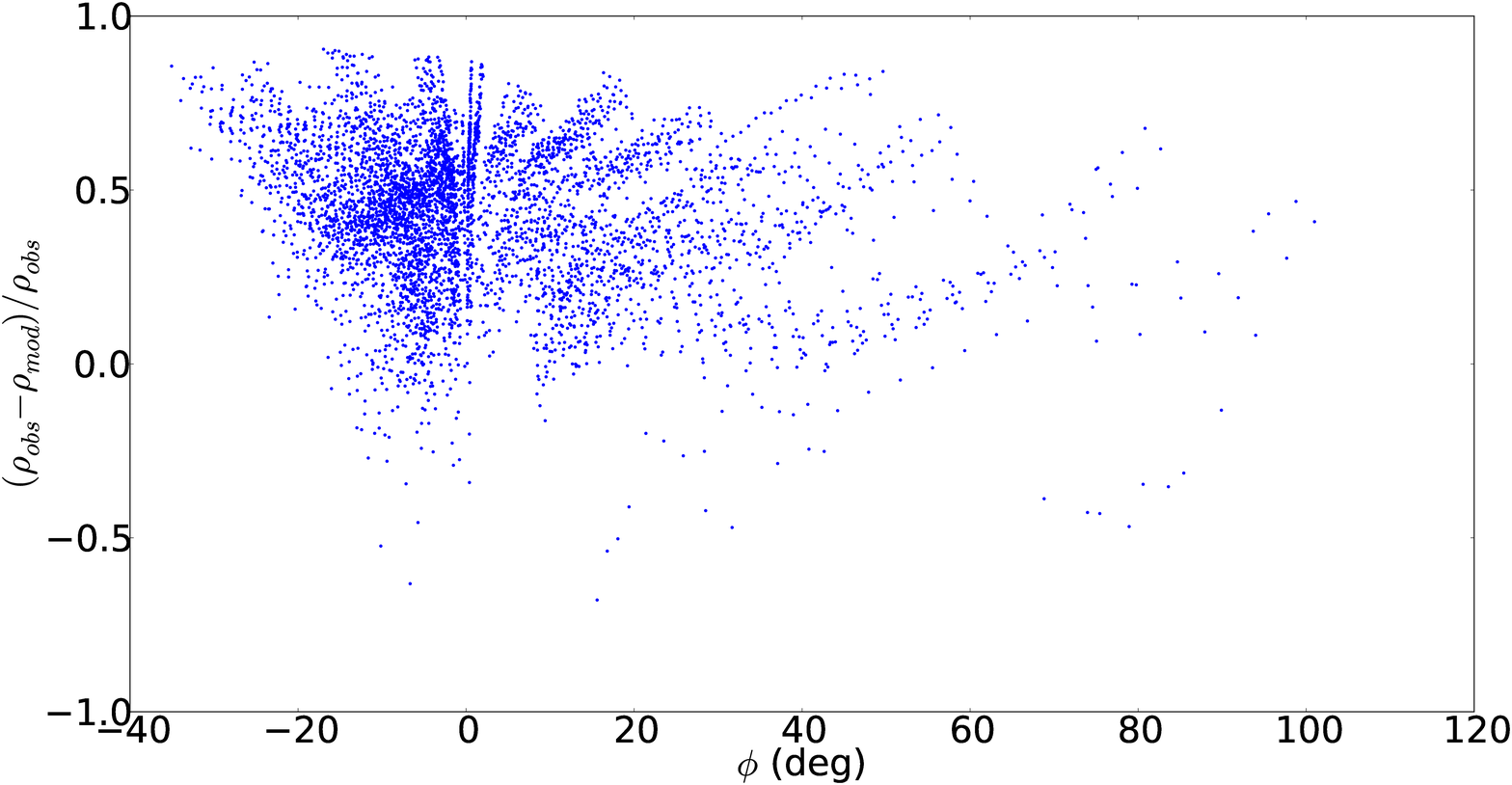}\\
\vspace{1cm}
\includegraphics[width=9cm]{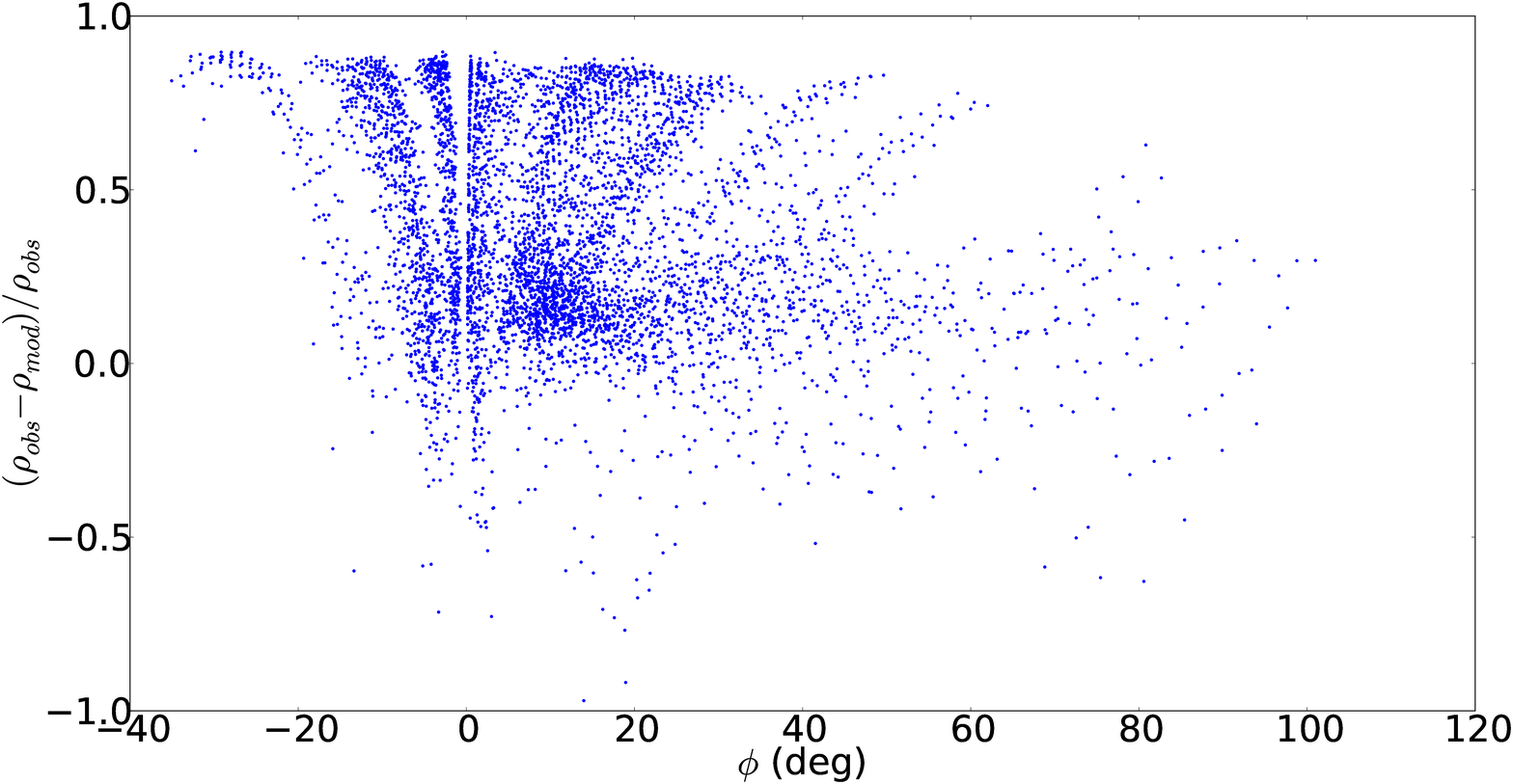}
\caption{Residuals of $\rho $ with respect to the best fit for the data extracted from the star counts of
F8-G5V stars in the SDSS. $R>7.5$ kpc, $|z|<3.5$ kpc. Top: $z>0$. Botton: $z<0$.}
\label{Fig:err_theta}
\end{figure}

\begin{table*}
\caption{Disc parameters of the best fit (see text). Note that there are nine independent parameters
in the fit; the last six parameters depend on the previous ones. The constraint 
$|\phi |\le 30^\circ $ explores the region where the warp amplitude is very low. The metallicity gradient
is modelled following Eq. (\ref{gradmet}).}
\begin{center}
\begin{tabular}{cccc}
\textrm{parameter} & \textrm{All data, no grad. metal.} & $|\phi |\le 30^\circ $, no grad. metal. & \textrm{All data, with grad. metal.}  \\
\hline
$h_{\rm r, thin}$ (kpc) & $2.0^{+0.3}_{-0.4}$  & 2.1  & 2.0  \\
$h_{\rm z, thin}(R_\odot)$ (kpc) & $0.24^{+0.12}_{-0.01}$  & 0.28  & 0.21  \\
$h_{\rm r, thick}$ (kpc) & $2.5^{+1.2}_{-0.3}$  & 2.5  & 2.7  \\
$h_{\rm z, thick}(R_\odot)$ (kpc) & $0.71^{+0.22}_{-0.02}$  & 0.60  & 0.55 \\ \hline
$k_{\rm 1, thin}$ (kpc$^{-1}$) & -0.037  & 0.090 & -0.190 \\
$k_{\rm 2, thin}$ (kpc$^{-2}$) & 0.052  & 0.043 & 0.070 \\
$k_{\rm 1, thick}$ (kpc$^{-1}$) & 0.021  & 1.448 & 0.000 \\
$k_{\rm 2, thick}$ (kpc$^{-2}$) & 0.006  & -0.055 & 0.010 \\
$R_{\rm ft}$ (kpc) & $6.9^{+10.1}_{-1.9}$ & 15.8 & 6.9 \\ \hline
$h_{\rm z, thin}(15\,{\rm kpc})/h_{\rm z, thin}(R_\odot)$   & 3.3$^{+1.8}_{-1.7}$    & 3.7   & 3.1 \\
$h_{\rm z, thin}(20\,{\rm kpc})/h_{\rm z, thin}(R_\odot)$   & 8.1$^{+5.4}_{-5.4}$    & 8.3   & 8.8 \\
$h_{\rm z, thin}(25\,{\rm kpc})/h_{\rm z, thin}(R_\odot)$   & 15.5$^{+10.8}_{-11.3}$  & 14.9  & 17.8 \\
$h_{\rm z, thick}(15\,{\rm kpc})/h_{\rm z, thick}(R_\odot)$ & 1.5$^{+4.8}_{-0.4}$   & 1.0  & 1.6 \\
$h_{\rm z, thick}(20\,{\rm kpc})/h_{\rm z, thin}(R_\odot)$  & 2.3$^{+12.9}_{-0.8}$  & 6.1   & 2.7 \\
$h_{\rm z, thick}(25\,{\rm kpc})/h_{\rm z, thick}(R_\odot)$ & 3.4$^{+25.4}_{-1.7}$   & 9.6  & 4.2 \\ \hline
\label{Tab:bestfit}
\end{tabular}
\end{center}
\end{table*}

We used the selected regions for the fit, excluding bins with a density lower than
$3\times 10^{-9}$ star pc$^{-3}$. For the fit of the scale lengths and scale 
heights ($h_{\rm r, thin}$,
$h_{\rm z, thin}(R_\odot)$, $h_{\rm r, thick}$, $h_{\rm z, thick}(R_{\rm ft})$), we used 
the regions with $|z|\le 3$ kpc, $R<15$ kpc, i.e., we neglect the flare; after we obtained these four parameters we fitted the rest of them for the flare in the regions with $1.5<|z({\rm kpc})|\le 3.5$ kpc, $7.5<R({\rm kpc})<30$. Within these ranges, the Galaxy is dominated by the disc rather than the halo star counts.

The parameters that produce the best weighted fit are given in Table \ref{Tab:bestfit}.
Figs. \ref{Fig:fitres} and \ref{Fig:err_theta} show how this model fits the data.
The error bars (1$\sigma $) of these parameters were derived using the method of Avni (1976) for four free
parameters (the first fit of the scales) and
five free parameters (the fit of the flare): $\Delta \chi^2=4.72$ and 5.89 respectively; 
where we normalized $\chi ^2$ such that 
$\chi _{\rm min}^2=N$ (number of data points) to take into account that the Poissonian
error is only a small part of the total error.

\begin{figure}
\vspace{1cm}
\centering
\includegraphics[width=8cm]{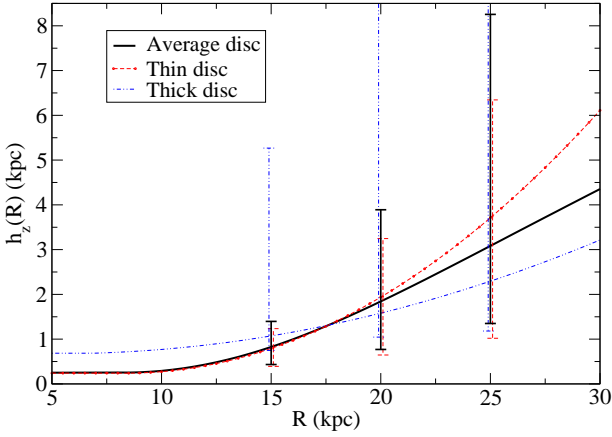} 
\caption{Scale height of the thin and thick discs according to our best fit.
The average is defined as the $-\left(\frac{d\,ln\,\rho _{\rm disc}(R,z=0)}{d|z|}\right)^{-1}$
and takes into account the increasing ratio of thick disc stars outwards.}
\label{Fig:flare}
\end{figure}

It can be observed how the flare becomes strong for high $R$ and $z$. In Fig. \ref{Fig:flare}, we plot the functional shape of $h_{\rm z, thin}(R)$ and $h_{\rm z, thick}(R)$ in this best-fit model.
The error bars for the flare amplitude are high, but they exclude the non-flared solution.
Note that the errors for the thin disc include the free variation of the thick disc to compensate for the
respective variations within those errors and vice versa. If we fixed one of the discs, the errors bars of
the other disc would be much lower than the case with its variation. When we combine both discs, we also expect
to reduce the error of the average disc (the errors of the average are also derived through the
$\chi ^2$ analysis), from which we conclude that we need a flare to fit our data, although distinguishing between thin and
thick disc component is only possible with low significance. 
It is clear from Fig. \ref{Fig:fitres} that an extrapolation of the
disc fitted at $R<15$ kpc without flare does not work at larger $R$: the shape of the density presents
a more complex form than the dashed line in the log-linear plot of Fig. \ref{Fig:fitres}.

\begin{figure}
\vspace{1cm}
\centering
\includegraphics[width=8cm]{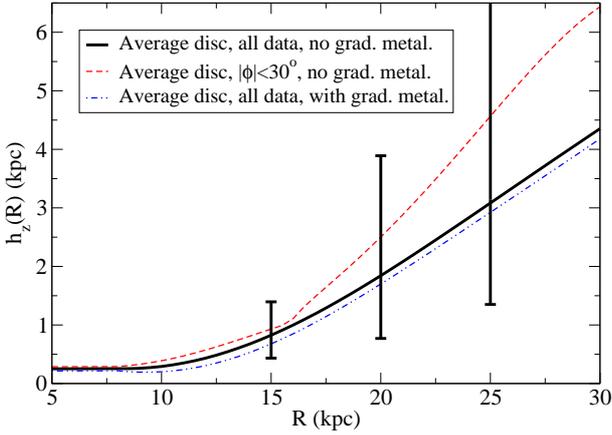} 
\caption{Scale height of the average defined as in Fig. \protect{\ref{Fig:flare}} for the best fits 
of Table \protect{\ref{Tab:bestfit}}.}
\label{Fig:flare2}
\end{figure}

We can see how the variation of the density with $z$ for
high $R$ becomes quite weak: green extends to almost all regions with high $R$ in Fig. \ref{Fig:dens}, and the few blue areas (very low densities) represent the few fluctuations due possibly
to some errors in the extinction or warp presence
(because the lowest latitudes are more affected by these
fluctuations). This is precisely what we obtain in our fit, 
which indicates the existence of a conspicuous
flare: a very significant increase of the scale height of the disc both for the thin and thick discs.
In Fig. \ref{Fig:fitres}, we also observe the effect of the flare: at $R\approx 15$ kpc and high $z$ the
density becomes almost constant with little decrease with $R$ 
because of a combination of the exponential decrease with radius and the increase of the density at high $z$ due to the increase of the scaleheight.

Fig. \ref{Fig:err_theta} shows some residuals in the observed stellar 
density with respect to the best fit. If the northern warp effect were significant, it would produce an excess density at $60^\circ \lesssim \phi \lesssim 100 ^\circ $ at $z>0$ and a deficit of stars in the same azimuths at $z<0$, and we do not observe that.
In general, some deviations from null residuals may be due to different effects such as metallicity gradients, a different
contamination of the halo than expected (at high $|z|$), some irregularities in the distribution of the extinction,
some degree of lopsidedness, or a variation of $h_z$ with $\phi $ (L\'opez-Corredoira \& Betancort-Rijo 2009).

\section{Exploring the effects of the warp and metallicity gradients}
\label{.explor}

We carried out two numerical experiments to determine the effects of 
the warp and the metallicity gradients. Since 
there are large uncertainties for both, we do not expect to derive direct
conclusions from the results of the following best fits, but they should serve as an estimate of the typical variation in our parameters under these changes. 

\begin{description}

\item[Warp:] instead of the whole data set, we used only those with $|\phi |\le 30^\circ $, where the amplitude
of the warp is lowest. We fitted $\rho (R,z)$ in the same way as before and the parameters obtained are those given in Table \ref{Tab:bestfit} and Fig. \ref{Fig:flare2}. 
They are compatible with the previous values, confirming our guess that the warp 
does not strongly change our results.

\item[Metallicity gradient:] we attributed to each star a Galactocentric distance $R'$, $\phi '$ and vertical position $z'$ corresponding to their coordinates and a distance $r'(m)$ given by
\begin{equation}
\label{gradmet}
r'(m)=10^{[m-M'+5]/5}
,\end{equation}\[
M'=4.8-0.4\Delta[Fe/H]
,\]\[
\Delta[Fe/H](R'[{\rm kpc}],z'[{\rm kpc}])=
\]\[
\left\{ \begin{array}{lcl}
      -0.078(R'-R_\odot)-0.220|z'| &  , R'\le f(z') \\
      -1 & , R'>f(z') \\
   \end{array}
   \right.
,\]\[
f(z')=R_\odot +12.82-2.82|z'|
.\]

As said in \S \ref{.method}, this stems from the variation of the 
absolute magnitude of F8V-G5V stars 
estimated by Siegel et al. (2002) for a variation of matallicity with respect to the Sun 
[$\Delta M_R\approx 0.4$ for $\Delta[Fe/H]=-1$ using $R-I=0.38$, which is the corresponding transformation from SDSS to Johnson filters (Jordi et al. 2006) of the
color of our population with average $(r-i)=0.13$ (Bilir et al. 2009); $\Delta M_g\approx \Delta M_R$ 
(Juri\'c et al. 2008, Fig. 3)], and considering a variation of metallicity from the combination of radial and vertical gradients of metallicity for the disc given by Rong et al. (2001) and Ak et al. (2007); assuming the
same gradient for thin and thick discs and that 
it remains constant for the farthest disc for a metallicity lower than the solar one.
This is probably an overestimation of $|\Delta[Fe/H]|$, which should be lower than unity at least for the
thin disc (Andreuzzi et al. 2011), but it serves as a limit of the strongest 
effect of this gradient.
Given that $\Delta[Fe/H](R',z')$ depends on the position and the position depends on this variation of metallicity, we carried out the calculation with an iterative process.

Then, we fitted $\rho (R',z')$ in the same way as before and the parameters obtained are
those given in Table \ref{Tab:bestfit} and Fig. \ref{Fig:flare2}. The results for the flare parameters
are totally compatible within the error bars with those obtained without taking into account any gradient
of metallicity.
As said in \S \ref{.method}, a lower metallicity in the farthest parts of the disc may overestimate
the distance of the stars (unfortunately, we do not know by how much since there are no 
accurate measurements of the metallicity of stars at those Galactocentric distances; we estimate an error of no more
than 20\% [\S \ref{.method}]), but our numerical experiment shows that 
the variation of the scale lengths and scale heights is slight and
the need of the flare is beyond these uncertainties with the metallicity.

\end{description}

\section{Comparison with other works}
\label{.compar}

The scale lengths and scale heights can be compared with those in other publications, although mainly for low $R$. Table \ref{Tab:compar} gives some of the values of the literature.
Jia et al. (2014, Table 1) reported many other values.
In general, the relative trends between thin and thick disc
are similar in all these papers and our results, but there
is some variation in the absolute scales that might arise because of different observed populations, different techniques of distance estimation, or different regions of application, apart, of course, from
possible systematic errors. Jia et al. (2014) have shown how the parameters, especially the 
scale height of the thin disc, depend on the absolute magnitude of the main-sequence stars used, indicating
that different populations have different velocity dispersions. The numbers of
Juri\'c et al. (2008) are somewhat higher than ours, possibly because they represent a range of smaller $R$, or possibly because
of their method of distance determination. Bilir et al. (2006) showed that  
the scale length depends on Galactic longitude. Here, we derived its average value but did
not examine any dependence on Galactic coordinates. 

\begin{table*}
\caption{Some values of the scale lengths and scale heights from the literature (units in kpc), 
derived either with SDSS (visible) or 2MASS (near infrared red).}
\begin{tabular}{ccccccc}
Reference & source & spatial range & $h_{\rm r, thin}$ & $h_{\rm z, thin}(R_\odot)$ & $h_{\rm r, thick}$ & $h_{\rm z, thick}(R_\odot)$ \\
\hline
L\'opez-Corredoira et al. (2002) & 2MASS & low Gal. latitudes & 3.3 & 0.28 & -- & -- \\ 
Cabrera-Lavers et al. (2005) & 2MASS & high Gal. latitudes & -- & 0.27 & -- & 1.1  \\
Bilir et al. (2006) & SDSS & interm. Gal. latitudes & 1.9 & 0.22 & -- & --  \\
Cabrera-Lavers et al. (2007) & 2MASS & interm. Gal. latitudes & -- & 0.19 & -- & 0.96  \\
Bilir et al. (2008) & SDSS & high Gal. latitutes & -- & 0.19 & -- & 0.63  \\
Juri\'c et al. (2008) & SDSS & $r<1.5$ kpc & 2.6 & 0.30 & 3.6 & 0.90  \\
Chang et al. (2011) & 2MASS & interm.-high Gal. latitudes & 3.7 & 0.36 & 5.0 & 1.0  \\
Polido et al. (2013) & 2MASS & whole sky & 2.1 & 0.21 & 3.0 & 0.64 \\
Jia et al. (2014) & SDSS+SCUSS & interm. Gal. latitude & -- & 0.20 & -- & 0.60 \\ 
\label{Tab:compar}
\end{tabular}
\end{table*}

The flare was previously observed by Alard (2000), 
L\'opez-Corredoira et al. (2002), Yusifov (2004), Momany et al. (2006), or Reyl\'e et al. (2009).
Polido et al. (2013) also introduced a flare in their model, but did no fit their parameters.
The values of the scale height at high $R$ from our fit (see Fig. \ref{Fig:flare}) are high: with
a thin disc scaleheight around 0.8 kpc at R=15 kpc, lower than the
extrapolation of L\'opez-Corredoira et al.
(2002) or that of Yusifov (2004), and
higher than the values of Alard (2000), Momany et al. (2006) and Reyl\'e et al. (2009). Between 20-25 kpc we derive a
thin disc scale height of 2-4 kpc, which is also higher than the values by
Momany et al. (2006) and Reyl\'e et al. (2009)  
(for the rest of the authors, there are no values at such 
high values of $R$).
For the thick disc, there are few or no studies to compare our studies with: 
there is a hint with unconclusive results  
by Cabrera-Lavers et al. (2007), which was constrained within $R<10$ kpc and
obtained the opposite sign in the increase of scale height for the solar neighbourhood;
it is interesting that the values of the flare in the thick disc 
that we obtained are those that are needed to explain the Monoceros ring in terms of 
Galactic structure (Hammersley \& L\'opez-Corredoira 2011).

Note that in our results at $R\gtrsim 17$ kpc 
the thin disc becomes ``thicker'' than the so-called thick disc.
This should motivate us to change the nomenclature: maybe instead of thin disc + thick disc
we should speak of disc 1 and disc 2. In any case, our model is just an exercise of fitting
stellar densities and, within the error bars we are unable to see which disc has a larger scale height
at large Galactocentric radius. We do not aim here to distinguish among different populations.
We see from our results that we need a flare to interpret the global density, but, with
the present analysis, we are not able to distinguish  the populations
that are flared at high $R$. We expect that at high $R$ there should be no significant difference between both discs
thicknesses and we have only one mixed component. An average disc as plotted in 
Fig. \ref{Fig:flare} represents this average old population of type F8V-G5V stars.
In this average disc, at high $R$ 
the thick disc (or better: disc 2) has a higher ratio of stars than at $R_\odot $:
while at $R_\odot $ it is 9\% of the stars of the thin disc (or better: disc 1), at $R=25$ kpc it is 50\% of the stars of the thin disc, because the scale length of the thick disc is larger than that of the thin disc, and consequently the fall-off of the density is slower.
 
The theoretical explanation of the observed features is beyond the scope of this paper.
The origin of the thick disc and its flare need to be
predicted by a model that aims to understand these observations. One of the hypotheses for the formation of thick discs is through minor mergers, which predicts a scale length of the thick disc larger than the scale length of the thin disc, a flare of increasing scale height in the thick disc, and a constant scale height for the stellar excess added by the merger (Qu et al.
2011). If we assumed that the mixture of the thin disc of the primary galaxy 
plus the stellar excess due to the accreted minor galaxy produces what we observe as the thick disc, our results would be fitted by those predictions. Nonetheless, our analysis is very rough, and without a necessary study of the populations we are unable to confirm this scenario.

\section{Discussion and conclusions}
\label{.discu}

Our method of deriving the 3D stellar distribution is quite straightforward, although it may contain some errors due mainly to inappropriate extinction estimate (some systematic error in the scales may be produced, but we do not expect it to exceed 18\%; see
\S \ref{.method}), metallicity gradients, or the effect of the warp. 
Even taking into account these factors, 
we have not observed features that suggested that the derived morphology
might be very different: the scales might change slightly, 
but the presence of the flare is unavoidable.

Our results show that the stellar density distribution of the outer disc (up to $R=30$ kpc) is well fitted by a component of thin+thick disc with flares (increasing scale height outwards). From our diagrams, it is clear that there is no a cut-off of the stellar component at $R=14-15$ kpc
as stated by Ruphy et al. (1996) or Minniti et al. (2011); we only examined off-plane regions 
($|b|\ge 8^\circ $) so we cannot judge what occurs in the in-plane regions, but from our results and by interpolating the results in the $z$-direction one can clearly conclude the reason why Ruphy et al. or Minniti et al. appreciated a significant drop-off of stars at $R=14-15$ kpc: the flare becomes important at those galactocentric distances, and consequently, the stars are distributed in a much wider range of heights, producing this apparent depletion of in-plane stars. Indeed, our Galactic disc does not present a cut-off there but the stars are spread in off-plane regions, even at $z$ of several kpc up to 
a Galactocentric distance of 15 scale lengths. Assuming that our fit is correct, for a constant luminosity function along the disc, the flux of the Milky Way seen observed face-on would follow a dependence
(from Eq. (\ref{rho}), neglecting the hole of the inner disc, which is totally insignificant for $R>R_\odot $)
\begin{equation}
F(R)\propto \int _{-\infty}^\infty dz\,\rho_{\rm disc}(R,z)
\end{equation}\[\,\,\,\,\,\,
\approx \rho _\odot \left[\exp\left(-\frac{R}{h_{\rm r, thin}}\right)+f_{\rm thick}\exp\left(-\frac{R}{h_{\rm r, thick}}\right)\right]
.\]
It is clear that in $F(R)$ there is no radial truncation in the explored range, and if this $F(R)$ did not represent our Galaxy, we would not derive as good a fit as we did. This does not mean that radial truncations are not possible in spiral galaxies: there are other galaxies in which it is observed (van der Kruit \& Searle 1981; Pohlen et al. 2000), but the Milky Way is not one of them.

The smoothness of the observed stellar distribution also suggests that there is a continuous structure
and not a combination of a Galactic disc plus some other substructure or extragalactic component. The
hypothesis of interpreting the Monoceros ring in terms of a tidal stream of a putative accreted dwarf galaxy
(Sollima et al. 2011; Conn et al. 2012; Meisner et al. 2012; Li et al. 2012)
is not only unnecessary (as stated by Momany et al. 2006; Hammersley \& L\'opez-Corredoira 2011; L\'opez-Corredoira et al. 2012), but appears to be quite inappropriate:
we see in Fig. \ref{Fig:dens} no structure overimposed on the Galactic disc.
Instead, the observed flare explains the overdensity in the Monoceros ring
observed in the SDSS fields (Hammersley \& L\'opez-Corredoira 2011).

Given these results, it would be interesting for future works if the dynamicists explained the existence of the observed flares in the Galactic disc, and further observational research on the spectroscopical
features of the disc stars at very high $R$ and high $z$ is necessary 
to know more about the origin and evolution of this component.

\begin{acknowledgements}
Thanks are given to A. Cabrera-Lavers, S. Zaggia and the anonymous referee for useful comments that helped us to improve this paper. MLC was supported by the grant AYA2012-33211 of the Spanish Ministry of Economy and Competitiveness (MINECO). Thanks are given to Astrid Peter (language editor of A\&A) for proof-reading of the text.

Funding for SDSS-III has been provided by the Alfred P. Sloan Foundation, the Participating Institutions, the National Science Foundation, and the U.S. Department of Energy Office of Science. The SDSS-III web site is http://www.sdss3.org/.
SDSS-III is managed by the Astrophysical Research Consortium for the Participating Institutions of the SDSS-III Collaboration including the University of Arizona, the Brazilian Participation Group, Brookhaven National Laboratory, Carnegie Mellon University, University of Florida, the French Participation Group, the German Participation Group, Harvard University, the Instituto de Astrofisica de Canarias, the Michigan State/Notre Dame/JINA Participation Group, Johns Hopkins University, Lawrence Berkeley National Laboratory, Max Planck Institute for Astrophysics, Max Planck Institute for Extraterrestrial Physics, New Mexico State University, New York University, Ohio State University, Pennsylvania State University, University of Portsmouth, Princeton University, the Spanish Participation Group, University of Tokyo, University of Utah, Vanderbilt University, University of Virginia, University of Washington, and Yale University. 
\end{acknowledgements}

\end{document}